\documentclass[conference]{IEEEtran}
\IEEEoverridecommandlockouts
\usepackage{cite}
\usepackage{amsmath,amssymb,amsfonts}
\usepackage{algorithmic}
\usepackage{graphicx}
\usepackage{textcomp}
\usepackage{textcmds}
\usepackage{subfigure}
\usepackage{xcolor}
\usepackage{url}
\def\BibTeX{{\rm B\kern-.05em{\sc i\kern-.025em b}\kern-.08em
    T\kern-.1667em\lower.7ex\hbox{E}\kern-.125emX}}
\begin{document}

\title{Microwave lymphedema assessment using deep learning with contour assisted backprojection}

\author{\IEEEauthorblockN{Yuyi Chang, Nithin Sugavanam, and Emre Ertin}
\IEEEauthorblockA{\textit{Department of Electrical and Computer Engineering} \\
\textit{The Ohio State University}\\
Columbus, OH 43210, USA \\
Email: chang.1560@osu.edu}
}

\maketitle

\begin{abstract}
We present a method for early detection of lymphatic fluid accumulation in lymphedema patients based on microwave imaging of the limb volume across an air gap. The proposed algorithm uses contour information of the imaged limb surface to approximate the wave propagation velocity locally to solve the eikonal equation for implementing the adjoint imaging operator. This modified backprojection algorithm results in focused imagery close to the limb surface where lymphatic fluid accumulation presents itself. Next, a deep neural network based on U-Net architecture is employed to identify the location and extent of the lymphatic fluid. Simulation studies with various upper and lower arm profiles compare the focusing performance of the proposed contour assisted backprojection imaging with the baseline imaging approach that assumes homogeneous media. The empirical results of the simulation experiments show that the proposed imaging method significantly improves the ability of the deepnet model to identify the location and the volume of the excess fluid in the limb. 
\end{abstract}


\section{Introduction}\label{intro}

Lymphedema is a chronic and progressive disease caused by impaired lymphatic drainage. 
This disease is a disabling, painful, and disfiguring complication of nodal therapy for various cancer types, affecting roughly 2-3 million people in the United States alone, representing 10-30\% of cancer survivors \cite{lym-rasmussen2010human}. 
The standard care of limb volume assessment through perometry has limited utility for early detection, as limb volume changes occur late in disease progression, long after the liquid content of subcutaneous tissues begins to increase. 
Existing medical imaging modalities like Magnetic Resonance Imaging (MRI) and X-ray can be used to detect abnormalities in the tissue \cite{lym-rasmussen2010human}, but the associated cost hinders the accessibility in early-stage detection. 
Microwave bio-radar sensing is an emerging non-invasive and safe biosensing modality that can assess the dielectric properties of subcutaneous tissues sensitive to changes in liquid content.
These sensors are low cost, easy to deploy, and applicable for both home and clinical care settings, making them a viable solution to enable lymphedema early detection and intervention.

% \subsection{Related works}

% microwave biosensing
\textbf{Microwave imaging of biological tissues:} Multiple studies have investigated the use of radiofrequency (RF) based sensors to detect stroke, detect brain tumor, and screen breast cancer \cite{semenov2008microwave,  hopfer2017electromagnetic, bindu2006active, mobashsher2016portable} and established the efficacy of this modality in detecting permittivity changes.
Existing works commonly assume that the antenna is well-matched to the skin layer. This is achieved by optimizing the antenna's impedance to match the tissue profile or immersing the imaged object in a liquid matching medium. 
Yet, this immersion setup complicates the measurement process and, therefore, is not suitable for frequent diagnostic measurements.
Microwave imaging for lymphedema diagnostics has yet to receive much attention, with the only known study \cite{russo2021preliminary} focusing on an antenna design perspective. 

% deep learning in RF based biosensing
\textbf{Deep learning based microwave imaging: } Deep learning has achieved state-of-the-art performances in many imaging tasks within medical and natural image domains \cite{zhou2021review}. 
Deep learning-based methods facilitate the reconstruction of backscattered signals and identify abnormal tissues in biosensing scenarios \cite{shao2020microwave, gerazov2017deep}. 
These deep net applications are focused on narrow-band systems based on simulations with idealized radiators.  

\textbf{Dielectric property inversions: } Inferring the dielectric properties of internal tissues from microwave measurements is an ill-posed inverse problem \cite{winters2006estimation, civek2022bayesian}. 
The ill-conditioning of the forward operator worsens if an air gap exists between the antenna elements and the skin surface. 
This leads to an abrupt change in relative permittivity from the air-skin interface and, consequently, a large reflection coefficient compared to returns from internal tissues. 
Additionally,  the transfer function of the antenna elements needs to be calibrated in situ, as it includes unknown detuning effects when placed in close proximity of the body,  increasing the number of parameters that need to be estimated\cite{chang2023removing, civek2022bayesian}.

% \subsection{Contributions}

\begin{figure*}[t]
    \centering
    \includegraphics[width=\textwidth]{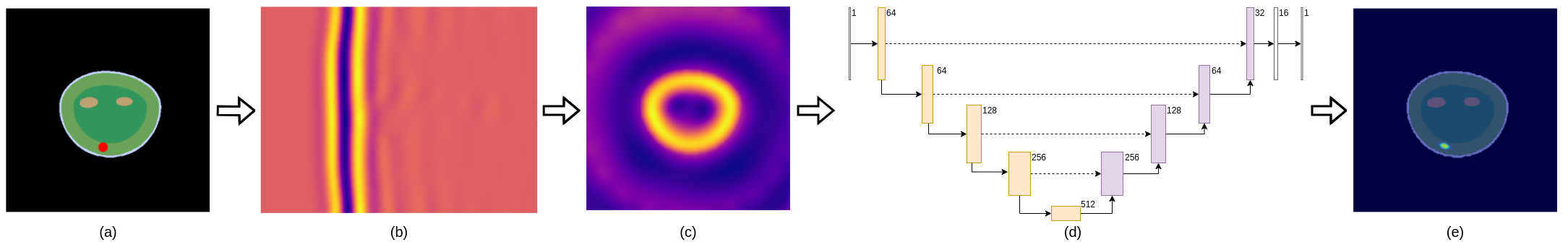}
    \caption{Illustration of our methodology: (a) a limb profile that is being assessed for lymphedema, with accumulated body fluid denoted in red and is unknown to the model; (b) tomographic data $P_i(t), i= 1, 2, ..., M$ collected/simulated based on the limb profile; (c) Contour guided backprojection image $I(\Vec{r})$; (d) deep learning U-Net model for localizing lymphedema; (e) model prediction of lymphatic fluid location superimposed on the limb profile.}
    \label{fig:model_sch}
\end{figure*}

We summarize our main contributions as follows
\begin{enumerate}
    \item Rather than requiring a matching medium with controlled dielectric properties, we develop a contour-guided backprojection method for imaging biological tissues across an air gap and validate using simulation experiments.
    \item We demonstrate how to use the contour information to create an approximate velocity map and employ a time-domain backprojection technique tailored for heterogeneous media, which leads to focused imagery close to the limb surface.
    \item We pair a deep learning based detector with the proposed counter-guided imaging method to detect the presence of excess body fluid within the limb. 
\end{enumerate}

This paper focuses on identifying lymphedema, but the methodology is adaptable to any detection/segmentation problems presented in microwave biosensing (e.g., finding tumors/stroke), in which the cancerous/abnormal tissues have different EM permittivity profiles compared to other tissues. 

\section{Counter guided time-domain backprojection algorithm}

We consider a circular radar array with $M$ individual elements that operates in a monostatic transmit and receive setting sensing the spatial domain $\Omega$. 
Each transceiver element comprising of a transmitter (TX) and a receiver (RX) element is spaced at a known distance $d$,  taking $M$ tomographic measurements around the limb equally spaced covering 360°.
The TX-RX pairs operate sequentially at individual angles in a time division multiplexing fashion. 
We further assume the surface contour of the limb is available from a secondary modality to provide an occupancy mask to guide the backprojection algorithm. 
In the following, we first review traditional time-domain backprojection algorithm in a homogeneous medium based on the Time-of-Flight (ToF) calculations. 
Next, we present a contour guided locally interpolated (CGLI) backprojection approach that takes into account a spatially varying EM propagation speed based on tissue profile prior.

% \subsection{Pointwise time-of-flight backprojection in homogeneous medium}
\subsection{Review of time-domain backprojection in homogeneous medium}\label{sec:simple_bp}

We consider a 2D spatial grid $\Omega$ containing $M$ radar elements. 
Each element, indexed by $i = 1, ..., M$, is equipped with a transmitter TX at $\Vec{T}_i \in \Omega$ and a receiver RX at $\Vec{R}_i \in \Omega$. 
For any point $\Vec{r} = (x, y) \in \Omega$, the travel time to $i^{th}$ radar element $t_i$ can be calculated by ToF. 

\begin{align}
\begin{split}
    t_i(\Vec{r}) &=  t_{T_i}(\Vec{r}) + t_{R_{i}}(\Vec{r}) \\
        % &= \frac{1}{v} \left( l_{TX_i} + l_{RX_i} \right) \\
        &= \frac{1}{v} \left(||\Vec{T}_i - \Vec{r}||_2 + ||\Vec{R}_i - \Vec{r}||_2 \right).
\end{split}
\label{eqn:tof}
\end{align}
where $v = c / \sqrt{\epsilon}$ is the wave propagation speed determined by permittivity $\epsilon$ of the medium with respect to the speed of light in a vacuum $c$.

The radar system produces a series of tomographic measurements $P(t) = \{ P_1(t), P_2(t), ..., P_M(t)\}$. 
We sample the backscattered received signal $P_i^S[m] = P_i(m \cdot \Delta t)$ at constant sampling interval $\Delta t$ for $m = 0, 1, ...,M$.
Signal contribution from the $i^{th}$ transmit-receive pair $I_i(\Vec{r})$ can be determined by interpolating $P_i(t)$ between $t_{k}$ and $(t_k + \Delta t)$ where $t_{k} < t_i(\Vec{r}) \le t_k + \Delta t$:

\begin{align}
\begin{split}
    I_i(\Vec{r}) &=  P_i(t_k) \cdot \frac{t_i(\Vec{r}) - t_k}{\Delta t} \\
                 &+  P_i(t_k+\Delta t) \cdot \frac{(t_k + \Delta t) - t_i(\Vec{r})}{\Delta t}
                 \label{eqn:bp_interpolate}
\end{split}
\end{align}

The backprojected signal intensity, denoted as $I(\Vec{r})$, is determined by the superposition of signals from all $M$ transmit-receive pairs. 

\begin{equation}
    I(\Vec{r}) = \sum_{i=1}^{M} I_i(\Vec{r}) 
    \label{eqn:bp_sum}
\end{equation}

\subsection{Contour guided locally interpolated backprojection}\label{sec:contour_bp}

We now examine the backprojection for the $i^{th}$ tx-rx pair $P_i(t)$ in a heterogeneous medium with spatially varying permittivity.
Wave propagation in this setup leads to multiple reflections due to the changing permittivity. 
More importantly, the wave velocity slows down due to increased permittivity in biological tissues resulting in increased spatial resolution when compared with free space propagation, thereby posing a challenge in identifying body fluid within internal tissues.

Given the $i^{th}$ radar element, we start by creating a travel time map   $\tau_{T_i}(\Vec{r})$ and $\tau_{R_i}(\Vec{r})$ that denotes the shortest propagation time  to each point $\Vec{r}=(x,y)$ in the scene originating from transmitter $T_i$ and receiver $R_i$ respectively. 

Inspired by the success in tracing the travel wave in geoscience applications, we employ an Anisotropic Locally Interpolated Fast Marching Method (ALI-FMM) \cite{alifmm-ludlam2023travel} to compute a 2D travel time in heterogeneous media.  The algorithm utilizes the velocity map of the propagation medium and iteratively calculates the total travel time from a starting point to all points on the spatial grid. 

We find the shortest travel time by solving the eikonal equation given by
\begin{equation}
    |\nabla\tau(\Vec{r})| = \frac{1}{v  \left( \Vec{r}, \nabla\tau/|\nabla\tau| \right)}
\end{equation}subject to boundary conditions:
\begin{align}
    \tau_{T_i}(\Vec{T}_i) &=  0 \quad
    \tau_{R_i}(\Vec{R}_i) =  0.
\end{align}
where $\tau(\Vec{r})$ is the shortest travel path, and $v(\Vec{r}, \nabla\tau/|\nabla\tau|)$ is the speed of the anisotropy media at point $\Vec{r}=(x,y)$.

In this work, we consider biological tissues are isotropic ~\cite{zhang2010imaging}. 
The wave velocity at point $\Vec{r}$ is independent of the traveling directions and can be reduced to $v(\Vec{r}, \nabla\tau/|\nabla\tau|) = v(\Vec{r})$.
The travel time calculation is iterative starting from a source point $\Vec{T}_i$ or $\Vec{R}_i$ and propagate to all $\Vec{r} \in \Omega$. 
The total travel time map $t_i(\Vec{r})$ is the sum of TX and RX propagation time where 
$t_i(\Vec{r}) = \tau_{T_i}(\Vec{r}) + \tau_{R_i}(\Vec{r})$.

We solve the eikonal equation iteratively, by following the steps in~\cite{alifmm-ludlam2023travel} and introduce three states for any point to keep track of the progress of traversal: \qq{known}, \qq{close}, and \qq{far}.
At the beginning, all points are initialized to \qq{far} state where none of the point has been visited. 
The points are updated to \qq{close} as soon as they have been traversed, indicating a middle state where a precise travel time has not been determined.
Traversal to a point is considered completed when it becomes \qq{known} state, meaning a travel time has been set after multiple optimization iterations. 
The iterations take a similar form as breadth-first search (BFS) search algorithm where the algorithm advances to the next closest spatial grid at each steps. 
However, in order to achieve a higher accuracy and to avoid  quantization error, the methodology employs a collection of 32 stencils to better approximate traveling wavefront using locally linear wavefronts. 
This results in additional search steps to select a stencil that achieves the best approximation. Again, the backprojected image $I(\Vec{r})$ can be obtained by aggregating individual single images $I_i(\Vec{r})$  following equation \ref{eqn:bp_interpolate}-\ref{eqn:bp_sum}.

\begin{figure}[htbp]
    \centering
    \subfigure[]{
        \includegraphics[width=0.46\columnwidth]{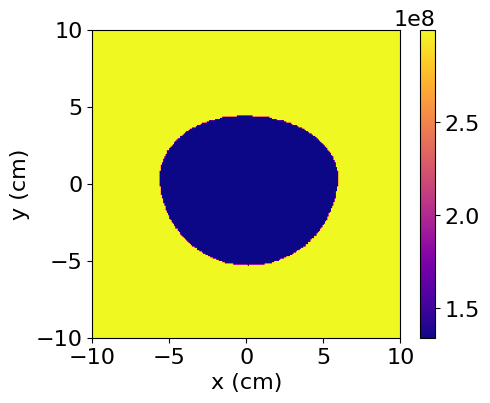}
        \label{fig:bp_step_vmap}
    }
    \subfigure[]{
        \includegraphics[width=0.46\columnwidth]{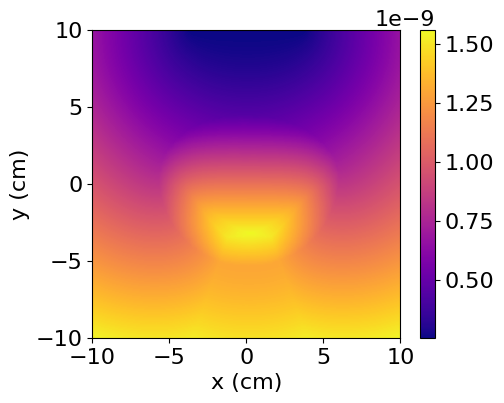}
        \label{fig:by_step_tau}
    }
    \subfigure[]{
        \includegraphics[width=0.46\columnwidth]{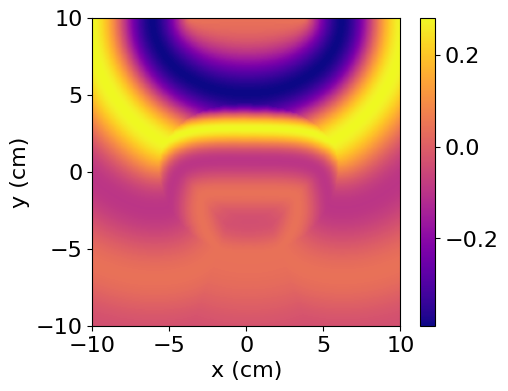}
        \label{fig:bp_step_single}
    }
    \subfigure[]{
        \includegraphics[width=0.46\columnwidth]{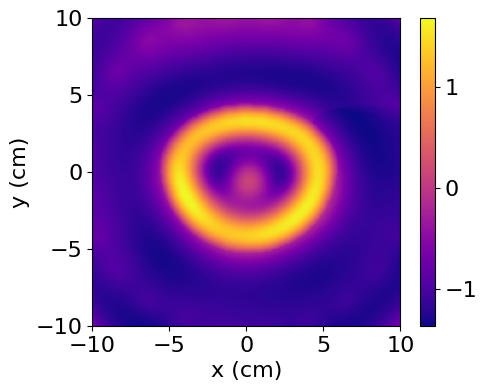}
        \label{fig:by_step_full}
    }

    % \subfigure[]{
    %     \includegraphics[width=0.45\columnwidth]{figs/bp_by_step/bp_ibp_full.png}
    %     \label{fig:by_step_ibp_full}
    % }
    
    \caption{Obtaining backprojected image $I(\Vec{r})$ from collected measurements $P(t)$ using CGLI backprojection method: (a) velocity map $v(\Vec{r})$ constructed based on limb occupancy; (b) two way travel time $t_i(\Vec{r}) = \tau_{T_i}(\Vec{r}) + \tau_{R_i}(\Vec{r})$; (c) backprojected image $I_i(\Vec{r})$ from a single TX/RX Pair. (d) full backprojected image $I(\Vec{r})$.}
      \label{fig:bp_by_step}
\end{figure}

Fig. \ref{fig:bp_by_step} illustrates individual steps to obtained a full backprojected image using CGLI method. 
A velocity map (Fig.~\ref{fig:bp_step_vmap}) is constructed based on arm occupancy information using an effective permittivity $\epsilon_{e}=5$. 
The velocity map enables travel time $\tau(\Vec{r})$ calculation (Fig.~\ref{fig:by_step_tau}).   
Fig.~\ref{fig:bp_step_single} shows single angle backprojected image $I_i(\Vec{r})$ by interpolating received signal $P_i(t)$ at the first radar element $i=1$.
The final backprojected image $I(\Vec{r})$ is obtained by repeating the previous steps for all 24 radar elements and aggregating the backprojection results (Fig.~\ref{fig:by_step_full}).

%%% U-Net
 \section{Deep learning model}

In this paper, we focus on a segmentation problem using a deep learning model designed to detect and localize lymphatic fluids. 
Using a backprojected image $I(\Vec{r})$ as the input, the model produces a mask that contains probabilistic measurements ranged from 0 to 1 indicating the likelihood of a given pixel containing body fluid. 
The detection of an elevated presence of body fluid is indicative of lymphedema. 

\begin{figure}[!htbp]
    \centering
    \includegraphics[width=\linewidth]{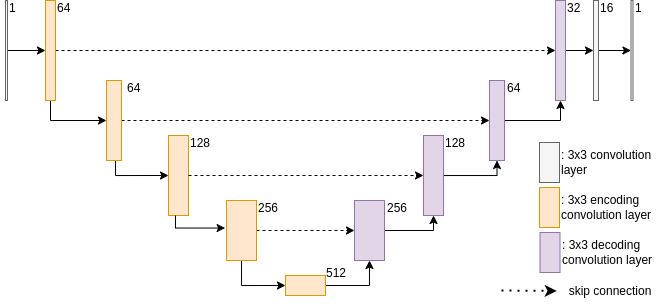}
    \caption{Illustration of deep learning U-Net model. The model contains multiple 3x3 convolutional blocks designed to extract image features at different dimensions. The model has an encoding (orange blocks) and a decoding path (purple blocks) that performs convolutional down/upsampling, respectively. The skip connections (dashed line) form direct connections between encoding and decoding paths to mitigate information loss.}
    \label{fig:unet}
\end{figure}

Our model is a deep convolutional neural network that is based on U-Net architecture \cite{unet-ronneberger2015u}. 
The U-Net model is first introduced for medical image segmentation tasks, and its design has demonstrated robust performances for other computer vision problems across different domains \cite{unet-ronneberger2015u, unetmwtumor-khoshdel2020full}. 
The architecture consists of an encoding path and a decoding path, forming a U-shape structure as shown in Fig. \ref{fig:unet}. 
Skip connections are placed between individual encoding and decoding levels in order to mitigate information loss during convolutional down/upsampling operations. 

We follow the configuration as in the original U-Net model \cite{unet-ronneberger2015u} to construct our model. 
Starting from a raw input image, the encoding path at every level condenses the input features to a higher-level description through multiple convolutional blocks. 
Each convolution block contains two sequentially operated 2D convolution, batch normalization and ReLU activation layers. 
At the end of each encoding level, the feature channel is doubled. 
An identical copy is cropped and passed through the skip connection to its corresponding decoder step, and a max pooling layer downsamples the original features by half for the next encoding step. 
At each decoding level on the decoding path, the skip connected feature tensor is concatenated with features from previous step. 
The combined features are upsampled by a convolutional layer and processed by the convolutional blocks following the same setup as in the encoding path. 

A final convolutional block contains a sigmoid layer, and is used to produce probability mask for localizing lymphatic fluid. 
Under binary segmentation formulation, the final sigmoid layer is typically combined into loss function for improved numerical stability. 
We further employ residual connections \cite{he2016deep} on encoding path to promote network's ability to identifying underlining body fluid. 
This results a hybrid deep learning model based on U-Net architecture \cite{unet-ronneberger2015u} for improved segmentation capacity. 

%%% experienments
\section{Experiments}

\subsection{Datasets}

We constructed upper extremity phantoms based on measurements from typical healthy adults. 
Ten base phantoms were generated, each of the incorporating layers representing skin, fat, muscle and bone tissues.
A cylindrical shape body fluid object is introduced within the fat layer mimicking the presence of lymphedema. 
The severity of lymphedema was controlled by varying the radius of the body fluid object. 

EM backscattered data were obtained through FDTD simulations using gprMax software \cite{warren2016gprmax}.
A tomography scan comprised samples taken at 24 uniformly spaced angles with a 20 x 20 cm domain. 
The ground truth segmentation masks were derived from tissue profiles as used in the simulation (e.g., Fig.~\ref{fig:model_sch}(a)). 
The simulation utilized a built-in bow-tie antenna, operating at 1.5 GHz with an approximate 1 GHz bandwidth. 
Calibration to simulated data was performed involving time axis correction using a set of size varying PEC cylinders, and free space calibration to enable coherently subtraction of antenna effects. 

Our EM dataset contain 676 samples which corresponds to 16224 individual simulations. 
We performed a random split of the dataset based on phantom profiles into training, validation and testing datasets. 
The distribution of samples is as follows: 520 samples for training, 52 samples for validation, and 104 samples for testing. 

EM simulations are inherently CPU intensive and time consuming tasks, making it crucial to accelerate data generations.  
We leverage GPU accelerated computing by offloading CPU tasks to multiple GPUs. 
A load balancing application was built to support executing a simulation task pool across multiple GPUs. 
This resulted approximately an 80x acceleration compared to running solely on CPU. 

\subsection{Implementation details}

Backprojected images take the shape of 256x256 with a single pixel representing roughly a 0.78x0.78 mm grid in spatial domain. 
A min/max scaling is used to scale input images to between 0 to 1 to ensure our received signals are always up to scale. 
Ranndom rotation and flipping are used to augment the training data during training. 

PyTorch framework is used for training and testing lymphedema segmentation. 
The training and testing are performed on a single Nvidia Quadro RTX 6000 GPU and Intel Xeon CPU server.
Our U-Net model contains five encoding/decoding steps, compressing the input feature up to 512 channels as shown in Fig.~\ref{fig:unet}. 
We use pretrained ResNet-34 \cite{he2016deep} as the backbone for U-Net encoders. 
The model takes single channel backprojected image as input and produces a single channel output. 
A sigmoid layer is applied the model output to produce probability map of body fluid.
Binary cross entropy (BCE) loss is used for training between the output and ground truth segmentation masks.
The model is trained 200 epochs, and the model weights that achieve the lowest validation loss is used for further evaluation. 

\subsection{Evaluation metrics}

We evaluate our methodology by comparing segmentation performance to the non-contour assisted backprojection algorithm (Section \ref{sec:simple_bp}) using our test dataset. 
In addition to BCE test losses, we analyze the sensitivity of the models based on probability of detection $P_{D}$ and probability of false alarm $P_{FA}$ from receiver operating characteristic (ROC) curves. 
% The ideal threshold setting for segmenting lymphatic fluid is determined by fixing a $P_{FA}$. 
We further report the model F1 and the intersection over union (IoU) scores at a threshold set by a desirable $P_{FA}$:

\begin{equation}
    F_1 = \frac{TP}{TP + 0.5 \left( FP + FN\right)}
\end{equation}

\begin{equation}
    \text{IoU} = \frac{TP}{\left( TP + FP + FN\right)}.
\end{equation}

%%% results
\section{Results}

\subsection{Backprojection using different strategies}

\begin{figure}[htbp]
    \centering
    \subfigure[]{
        \includegraphics[width=0.4\columnwidth]{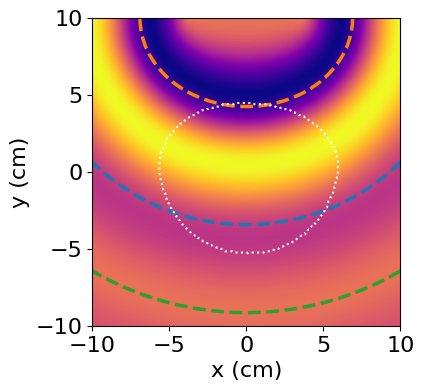}
        \label{fig:bp_compare_ibp}
    }
    \subfigure[]{
        \includegraphics[width=0.4\columnwidth]{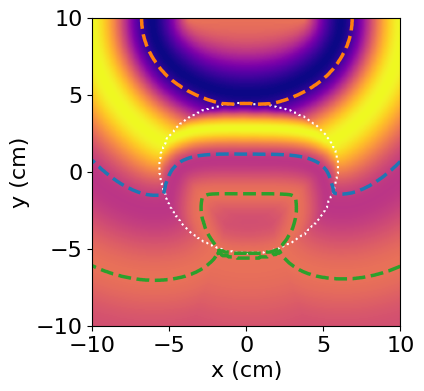}
        \label{fig:bp_compare_cgli}
    }
    \subfigure[]{
        \includegraphics[width=0.8\columnwidth]{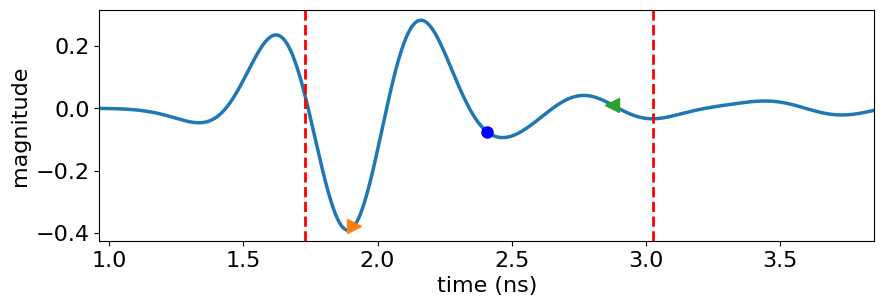}
        \label{fig:bp_compare_1D}
    }
    \caption{Equal travel time contours superimposed on the backprojected result using (a) ToF based method and (b) our CGLI backprojection methodology. (c) Original received signal $P_i(t)$ with markers indicating the corresponding location of equal travel time contours.}
    \label{fig:bp_compare}
\end{figure}

Fig.~\ref{fig:bp_compare} presents single angle backprojected image $I_i(\Vec{r})$ using ToF method (Fig.~\ref{fig:bp_compare_ibp}) ad our CGLI method (Fig.~\ref{fig:bp_compare_cgli}) with the phantom contour superimposed in white dashed lines. 
To better understand the effects of backprojecting in a heterogeneous medium, we draw three equal travel time contours to show the progression of backprojected signal.
We examine the relationships between the backprojected images and the original received signal (Fig. \ref{fig:bp_compare_1D}). 
The red dashed lines on each sides of the two images indicate the duration of backprojected time range. 
The orange and green markers indicate the duration during which the backprojected signal is within the phantom boundary, and an arbitrary time (blue marker) is picked to supplement the equal travel time contour. 
Between the two range markers, the propagation speed is reduced within the phantom as can be seen from the concave shape of the equal travel time contour in Fig.~\ref{fig:bp_compare_cgli}. 
In addition, the effects of shadowing were observed where a mirrored copies of the original signals appear in the backprojected image as evident in the green contour in Fig.~\ref{fig:bp_compare_cgli}. 
The shadowing occurs due to EM wave travel times are the same when penetrating the limb and travel along the air-skin interface before entering the internal tissues. 
The effect is predominant due to skin effects during EM wave propagation \cite{vander2006rf}. 
However, the effect does not contribute significantly to the backprojected results as the shadowing effects only occur in the lower regime of the lime when the backscattered signal intensity is low. 

\subsection{Lymphedema segmentation using deep learning models}

\begin{figure}[htbp]
    \centering
    \includegraphics[width=\linewidth]{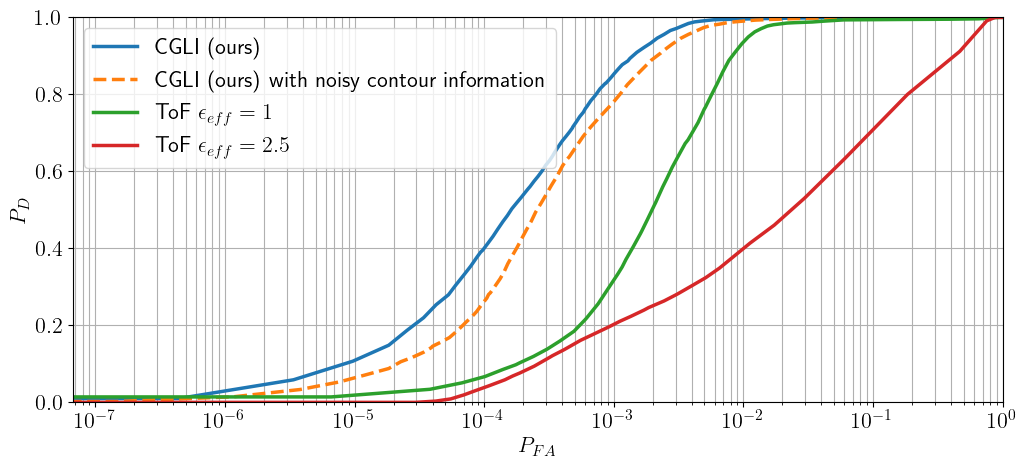}
    \caption{Reciver operating characteristics (ROC) of our method and baseline models showing probability of detection $P_D$ and probability of false alarm $P_{FA}$.}
    \label{fig:roc}
\end{figure}

We now present deep learning results for identifying lymphatic fluid within the tissues using a U-Net model.
Fig~\ref{fig:roc} presents the receiver operating characteristic (ROC) curves showing probability of detection $P_D$ with respect to false alarm probability $P_{FA}$.
The result shows our model achieves better detection accuracy compared to two baselines with different permittivity assumption under a fixed $P_{FA}$. 
At $P_{FA} = 10^{-3}$, the detection probability of our model is $P_D=0.852$ over baselines' $0.317$ and $0.202$.
We bias the model at $P_{FA} = 10^{-3}$ and report F1 and IoU scores in Table~\ref{tab:results}. 
Having the lowest testing BCE loss, our method achieves strong performance over other baseline models with F1 score of 0.565 and IoU of 0.393. 

To demonstrate the robustness of our model, we perturb the contour information by an Additive White Gaussian Noise (AWGN) simulating the uncertainty in contour measurements. 
We consider a \textit{worst case} scenario where the maximum position error of the contour can be up to +/- 1 mm, which is 200\% more than the typical error of commercial 3D scanner units.  
Our model continues to achieve strong performance over baseline methods with $P_{D} = 0.781$ and IoU of $0.360$. 

% test result summary
\begin{table}[htbp]
\caption{Summary of test results evaluated at $P_{FA}=0.001$}
\begin{center}
\begin{tabular}{|c|c|c|c|c|}
\hline
\textbf{Backprojection Method}  & \textbf{$\boldsymbol{P_D}$ $\uparrow$} & \textbf{F1 Score $\uparrow$} & \textbf{IoU $\uparrow$} \\ \hline
ToF Baseline ($\epsilon_{e} = 2.5$)       & 0.202 & 0.171 & 0.094  \\ \hline
ToF Baseline ($\epsilon_{e} = 1.0$)       & 0.317 & 0.255 & 0.146  \\ \hline
CGLI                                      & 0.852 & 0.565 & 0.393  \\ \hline
CGLI with noisy contour                   & 0.781 & 0.529 & 0.360  \\ \hline
\end{tabular}
\label{tab:results}
\end{center}
\end{table}

Our CGLI method enabled accurate detection of regions of lymphatic fluid accumulation. 
Using the contour information, we are able to create a focused imagery where the backprojected signals closely matched to the actual spatial location of the limb.
This is possible via time-domain CGLI backprojection algorithm tailored for heterogeneous media. 
Although the contour information alone is not sufficient to diagnose early stage lymphedema - a point we made in the opening section, the focused backprojected imagery encourages the deep learning detector to focus more on the changes in subcutaneous dielectric profiles happening near the limb surface. 
The output enables the lymphedema location appearing in the final segmentation mask to match the actual spatial location of the lymphedema, which improves the interpretability of any downstream tasks. 

\begin{figure}[htbp]
    \centering
    \subfigure[]{
        \includegraphics[width=0.97\columnwidth]{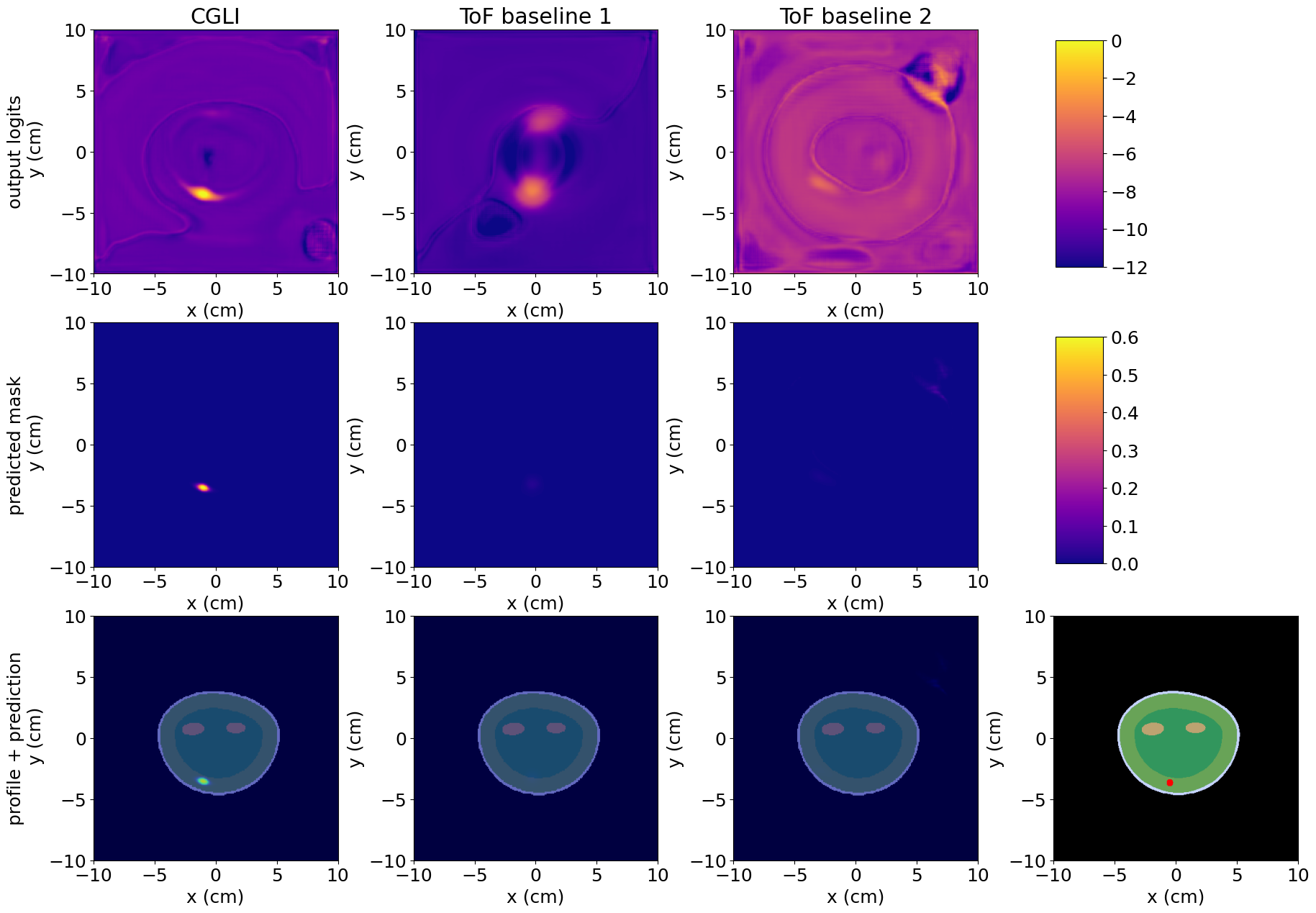}
        \label{fig:results_a}
    }
    \hfil
    \subfigure[]{
        \includegraphics[width=0.97\columnwidth]{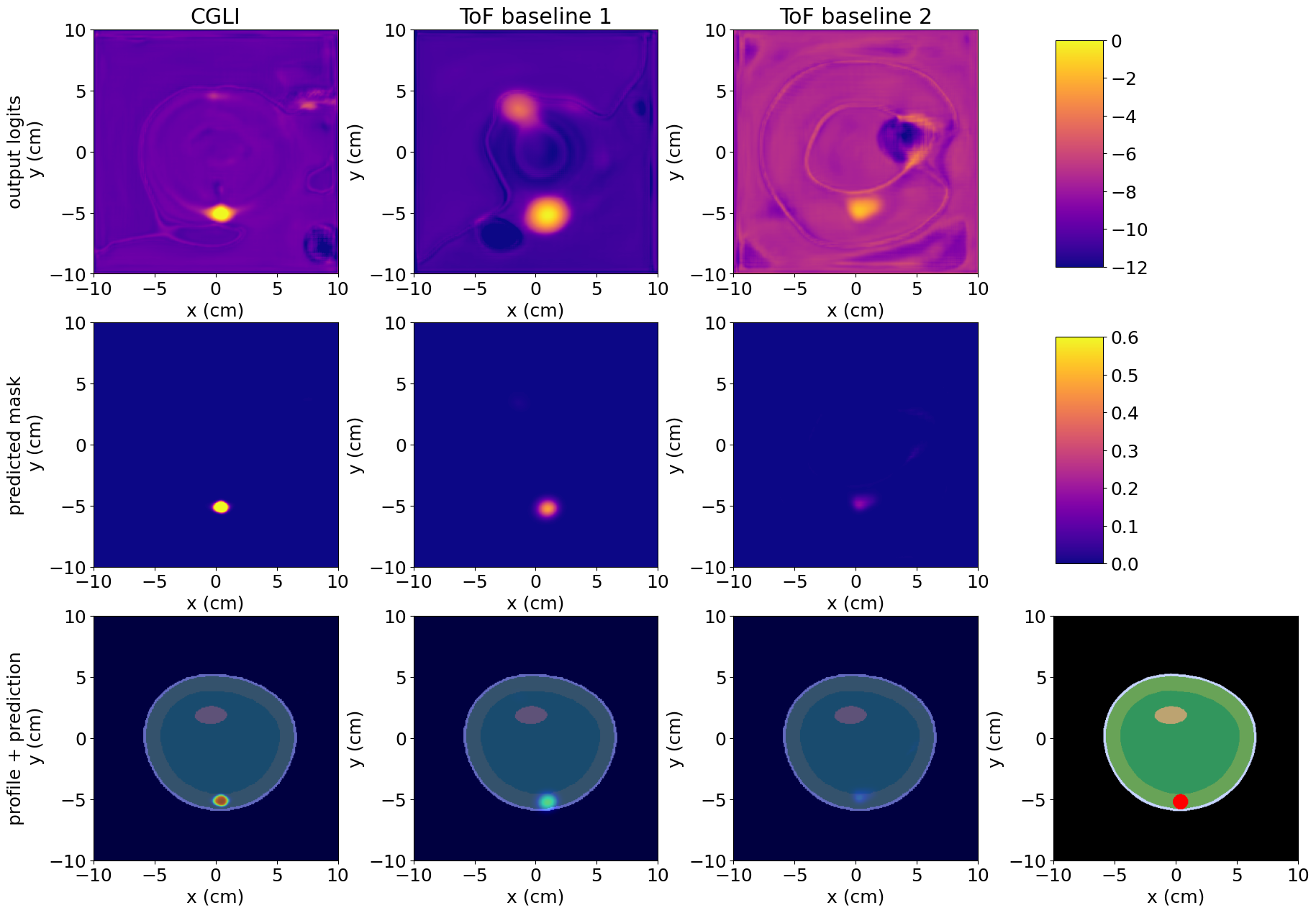}
        \label{fig:results_b}
    }
    \caption{Individual test samples comparing our method to baseline methods (left to right columns). On each subplot, intermediate model outputs including logits (row 1) and prediction probability map (row 2) are presented with assessment output superimposed on the ground truth profile (row 3). Our model shows confident and accurate predictions on (a) a lower limb with mild lymphedema $r=2.5$mm, and (b) a upper limb showing severe lymphedema $r=6$mm compared to baseline methods.}
    \label{fig:results}
\end{figure}

In the following we use two of the individual test samples to demonstrate the effectiveness of the proposed method over  the alternative baseline methods discussed above in Fig.~\ref{fig:results}.
The first test sample shows a lower limb with lymphatic fluid radius $r=2.5 $ mm, representing a mild state of lymphedema where a \textit{just-in-time} intervention is necessary (Fig.~\ref{fig:results_a}). 
The second test sample comes from a upper limb profile with a severe state of lymphedema at radius $r=6 $ mm (Fig.~\ref{fig:results_b}).
From top to bottom, we present 1) logit outputs from deep learning models, 2) predicted probability mask after the sigmoid of the final layer, and 3) predicted lymphedema location superimposed on the ground truth profile. 
Compared to baseline method with either permittivity assumption, our method precisely locate the location of lymphatic fluid on the phantom with less false positive.
With contour guided methodology, the target location is already aligned with actual geometry, so no additional calibration needs to be learned by the U-Net to establish the transformation between the lymphatic fluid location in microwave imaging domain to spatial domain.

%%% conclusion
\section{Conclusion}

In this paper, we presented a contour based backprojection imaging method paired with a  deep learning network that can localize excess liquid in subcutaneous tissues caused by lymphedema.
To enable imaging across an air gap, we use prior information about the surface contour of the limb to backproject backscatter signals to spatial domain accounting different propagation speed inside and outside the limb profile.  
The detection task of lymphatic fluid is solved using a deep learning segmentation U-Net model, which takes the backprojected image and produces a probability map where the fluid are likely to be present on the backprojected image. 
We evaluated our methodology on a synthetic microwave imaging dataset and compared our method to various baseline methods.
In future work, we plan to 1) develop a model that can infer, segment and account for further subcutaneous structures like fat tissue, 2) evaluate our method on measured data collected on phantoms.
We make the model and the imaging code publicly available at \url{https://github.com/SENSE-Lab-OSU/microwave_lymphedema_assessment}. 

\section*{Acknowledgment}

This work is partly supported by NSF Grant CBET-2037398 and NIH Grant P41EB028242.

\bibliographystyle{IEEEtran} 
\bibliography{bib}

\begin{thebibliography}{10}
\providecommand{\url}[1]{#1}
\csname url@samestyle\endcsname
\providecommand{\newblock}{\relax}
\providecommand{\bibinfo}[2]{#2}
\providecommand{\BIBentrySTDinterwordspacing}{\spaceskip=0pt\relax}
\providecommand{\BIBentryALTinterwordstretchfactor}{4}
\providecommand{\BIBentryALTinterwordspacing}{\spaceskip=\fontdimen2\font plus
\BIBentryALTinterwordstretchfactor\fontdimen3\font minus \fontdimen4\font\relax}
\providecommand{\BIBforeignlanguage}[2]{{%
\expandafter\ifx\csname l@#1\endcsname\relax
\typeout{** WARNING: IEEEtran.bst: No hyphenation pattern has been}%
\typeout{** loaded for the language `#1'. Using the pattern for}%
\typeout{** the default language instead.}%
\else
\language=\csname l@#1\endcsname
\fi
#2}}
\providecommand{\BIBdecl}{\relax}
\BIBdecl

\bibitem{lym-rasmussen2010human}
J.~C. Rasmussen, I.-C. Tan, M.~V. Marshall, K.~E. Adams, S.~Kwon, C.~E. Fife, E.~A. Maus, L.~A. Smith, K.~R. Covington, and E.~M. Sevick-Muraca, ``Human lymphatic architecture and dynamic transport imaged using near-infrared fluorescence,'' \emph{Translational Oncology}, vol.~3, no.~6, pp. 362--IN7, 2010.

\bibitem{semenov2008microwave}
S.~Y. Semenov, D.~R. Corfield \emph{et~al.}, ``Microwave tomography for brain imaging: Feasibility assessment for stroke detection,'' \emph{International Journal of Antennas and Propagation}, vol. 2008, 2008.

\bibitem{hopfer2017electromagnetic}
M.~Hopfer, R.~Planas, A.~Hamidipour, T.~Henriksson, and S.~Semenov, ``Electromagnetic tomography for detection, differentiation, and monitoring of brain stroke: A virtual data and human head phantom study.'' \emph{IEEE Antennas and Propagation Magazine}, vol.~59, no.~5, pp. 86--97, 2017.

\bibitem{bindu2006active}
G.~N. Bindu, S.~Abraham, A.~Lonappan, V.~Thomas, C.~Aanandan, and K.~Mathew, ``Active microwave imaging for breast cancer detection,'' \emph{Progress In Electromagnetics Research}, vol.~58, pp. 149--169, 2006.

\bibitem{mobashsher2016portable}
A.~T. Mobashsher, A.~Mahmoud, and A.~Abbosh, ``Portable wideband microwave imaging system for intracranial hemorrhage detection using improved back-projection algorithm with model of effective head permittivity,'' \emph{Scientific reports}, vol.~6, no.~1, p. 20459, 2016.

\bibitem{russo2021preliminary}
D.~Russo, ``Preliminary investigation on microwave sensing for early detection of lymphoedema,'' Ph.D. dissertation, Politecnico di Torino, 2021.

\bibitem{zhou2021review}
S.~K. Zhou, H.~Greenspan, C.~Davatzikos, J.~S. Duncan, B.~Van~Ginneken, A.~Madabhushi, J.~L. Prince, D.~Rueckert, and R.~M. Summers, ``A review of deep learning in medical imaging: Imaging traits, technology trends, case studies with progress highlights, and future promises,'' \emph{Proceedings of the IEEE}, vol. 109, no.~5, pp. 820--838, 2021.

\bibitem{shao2020microwave}
W.~Shao and Y.~Du, ``Microwave imaging by deep learning network: Feasibility and training method,'' \emph{IEEE transactions on antennas and propagation}, vol.~68, no.~7, pp. 5626--5635, 2020.

\bibitem{gerazov2017deep}
B.~Gerazov and R.~C. Conceicao, ``Deep learning for tumour classification in homogeneous breast tissue in medical microwave imaging,'' in \emph{IEEE EUROCON 2017-17th International Conference on Smart Technologies}.\hskip 1em plus 0.5em minus 0.4em\relax IEEE, 2017, pp. 564--569.

\bibitem{winters2006estimation}
D.~W. Winters, E.~J. Bond, B.~D. Van~Veen, and S.~C. Hagness, ``Estimation of the frequency-dependent average dielectric properties of breast tissue using a time-domain inverse scattering technique,'' \emph{IEEE Transactions on Antennas and Propagation}, vol.~54, no.~11, pp. 3517--3528, 2006.

\bibitem{civek2022bayesian}
B.~C. Civek and E.~Ertin, ``Bayesian sparse blind deconvolution using {MCMC} methods based on normal-inverse-gamma prior,'' \emph{IEEE Transactions on Signal Processing}, vol.~70, pp. 1256--1269, 2022.

\bibitem{chang2023removing}
Y.~Chang, N.~Sugavanam, and E.~Ertin, ``Removing antenna effects using an invertible neural network for improved estimation of multilayered tissue profiles using {UWB} radar,'' in \emph{2023 IEEE USNC-URSI Radio Science Meeting (Joint with AP-S Symposium)}.\hskip 1em plus 0.5em minus 0.4em\relax IEEE, 2023, pp. 53--54.

\bibitem{alifmm-ludlam2023travel}
J.~Ludlam, K.~Tant, V.~Dolean, and A.~Curtis, ``Travel times and ray paths for acoustic and elastic waves in generally anisotropic media,'' \emph{Journal of Computational Physics}, vol. 494, p. 112500, 2023.

\bibitem{zhang2010imaging}
X.~Zhang, S.~Zhu, and B.~He, ``Imaging electric properties of biological tissues by {RF} field mapping in {MRI},'' \emph{IEEE transactions on medical imaging}, vol.~29, no.~2, pp. 474--481, 2010.

\bibitem{unet-ronneberger2015u}
O.~Ronneberger, P.~Fischer, and T.~Brox, ``U-net: Convolutional networks for biomedical image segmentation,'' in \emph{Medical Image Computing and Computer-Assisted Intervention--MICCAI 2015: 18th International Conference, Munich, Germany, October 5-9, 2015, Proceedings, Part III 18}.\hskip 1em plus 0.5em minus 0.4em\relax Springer, 2015, pp. 234--241.

\bibitem{unetmwtumor-khoshdel2020full}
V.~Khoshdel, M.~Asefi, A.~Ashraf, and J.~LoVetri, ``Full 3d microwave breast imaging using a deep-learning technique,'' \emph{Journal of Imaging}, vol.~6, no.~8, p.~80, 2020.

\bibitem{he2016deep}
K.~He, X.~Zhang, S.~Ren, and J.~Sun, ``Deep residual learning for image recognition,'' in \emph{Proceedings of the IEEE conference on computer vision and pattern recognition}, 2016, pp. 770--778.

\bibitem{warren2016gprmax}
C.~Warren, A.~Giannopoulos, and I.~Giannakis, ``{gprMax}: Open source software to simulate electromagnetic wave propagation for ground penetrating radar,'' \emph{Computer Physics Communications}, vol. 209, pp. 163--170, 2016.

\bibitem{vander2006rf}
A.~Vander~Vorst, A.~Rosen, and Y.~Kotsuka, \emph{{RF}/microwave interaction with biological tissues}.\hskip 1em plus 0.5em minus 0.4em\relax John Wiley \& Sons, 2006.

\end{thebibliography}

\end{document}